\documentclass[a4paper,10pt]{article}
\usepackage{setspace}
\usepackage{ucs}
\usepackage[utf8]{inputenc}
\usepackage{amsmath}
\usepackage{amsfonts}
\usepackage{amssymb}
\usepackage{amsthm}
\usepackage{rotating}
\usepackage{xcolor}
\usepackage[english]{babel}
\usepackage[T1]{fontenc}
\usepackage{graphicx}
\usepackage{float}
\usepackage{hyperref}
\usepackage{cuted}
\usepackage{verbatim}
\usepackage{subfigure}
\usepackage{multirow}
\usepackage{bbold}
\usepackage{slashed}
\usepackage{indentfirst}
\usepackage{tcolorbox}
\usepackage{xcolor}
\usepackage{placeins}
\usepackage[a4paper, total={7in, 10in}]{geometry}

\usepackage{multicol}
\usepackage{fmtcount}
\usepackage{diagbox}
\usepackage{slashed}

\usepackage{placeins}

\title{\Large   \bf Magnetic field induced anomalous pion couplings}

\author{ Fabio L. Braghin$^{1}$,
Marcelo Loewe$^{2,3}$, Cristian Villavicencio$^{4}$ 
\\
{\normalsize $^{1}$ Instituto de F\'\i sica, Federal University of Goi\'as, Av. Esperan\c ca, s/n,
 74690-900, Goi\^ania, GO, Brazil}
\\
{\normalsize $^{2}$
Facultad de Ingenier\'{\i}a, Universidad San Sebasti\'an, Santiago, Chile}
\\
{\normalsize $^{3}$
Centre for Theoretical and Mathematical Physics, and Department of Physics, }
\\ 
{\normalsize
University of Cape Town,Rondebosch 7700, South Africa}
\\
{\normalsize $^4$
Centro de Ciencias Exactas $\&$ Departamento de Ciencias B\'asicas, Facultad de Ciencias, }
\\ {\normalsize
Universidad del B\'{\i}o-B\'{\i}o, Casilla 447, Chill\'an, Chile
}
}

\date{}

\begin{document}

\maketitle
\begin{abstract}
Effective pion-constituent quark couplings induced by 
relatively weak magnetic fields are calculated in the framework 
based in  Weinberg's large Nc Effective Field Theory. 
These couplings  (form factors)
vanish  in the vacuum. 
In particular,   single-pion
couplings to
a scalar and a vector 
constituent quark currents are investigated.
These couplings might  correspond to 
a fluctuation of the neutral  pion into a scalar quark-antiquark state  (meson)
and to a vector meson, respectively.
Some possible phenomenological implications are discussed.

\end{abstract}

\section{Introduction}

Strong magnetic fields,
of the order $10^{11}$ T$ - 10^{18}$ T, are expected to show up 
in dense stars, magnetars, and in peripheral relativistic heavy ion 
collisions (r.h.i.c.) besides probably having had 
effects in the early Universe
\cite{magnetars,early-universe,
review-B,review-B2,PPNP,Taya-2025}. 
They can be 
of the order of hadron masses, 
and therefore they 
can have a significant impact 
on hadron structure and dynamics
\cite{PPNP,magnetic-catalysis,Bruckamnn-etal,CME-VE,carlomagno-etal,LQCD1,LQCD2,D-Elia-etal,couplings,yukawaB2024}.
Taking into account magnetic field effects in virtual quark and antiquark
lines   usually  requires advanced
numerical efforts except if  particular limits are  considered.
The case of relatively weak magnetic fields, 
weak with respect to 
the pion or constituent quark masses,
 have been considered in 
several works for different systems \cite{weakB1,weakB2,Braghin2020,yukawaB2024,yukawaB-Miura-etal}.
In this limit, usually it is possible to address most part of the calculations analytically, 
with some needed reduced numerical effort.
 In fact, most recent estimations for the strength of magnetic fields
in h.i.c. or r.h.i.c. indicate that they should not be larger than in this range of relatively weak magnetic fields
\cite{Taya-2025,recent-rhic,sun-yan}.
Given the numerous difficulties in identifying the presence and the strength of magnetic fields,
 further effects and observables, theoretically predicted,
may contribute to these issues.
Furthermore, the effects of magnetic fields on hadron structure 
and dynamics can help to probe these systems that are very 
complicated and require enormous theoretical and experimental
efforts \cite{yukawaB2024,PPNP}.
In this context, in the present work, we address possible pion 
interactions induced by external magnetic fields.
Eventually these interactions may lead to sizable effects for observables
in r.h.i.c.

Pions are (quasi-)Goldstone bosons,
being their very small masses 
 as a consequence of 
explicit chiral symmetry breaking in 
QCD.
There are several possible pion-nucleon 
 couplings in the vacuum parameterized 
by  the different
Lorentz-flavor   structures \cite{gernot-fischer}.
The pseudoscalar and axial (single-) 
 pion couplings
represent a single pion attached to a corresponding fermion current
in such a way to preserve chiral symmetry, respectively, in 
the linear and non-linear realizations of chiral symmetry.
These couplings
are the most employed ones,
having a large support  from  phenomenology.
Usually, in the framework of the constituent quark model (CQM), all these couplings are understood as pion-nucleon   or pion - constituent quark
couplings \cite{Weinberg2010,plessas}. They  provide form factors that makes explicit the 
pseudoscalar (and axial) content of the pion.
There are also two-pion effective couplings   incorporated 
into chiral descriptions of low energy dynamics: the vector 
and the scalar pion couplings. 
All these pion couplings are expected to help the understanding of hadron (and nuclear) processes in terms of QCD-based description. They have been derived by starting with a quark-antiquark interaction due to one-gluon exchange in \cite{BraghinEFT} by considering standard analytical methods.
In Ref. \cite{yukawaB2024}
the authors have calculated corrections due to magnetic fields to the Yukawa potential between constituent quarks -as representative of nucleons.
The corresponding (pion) form factors have been 
investigated in the literature, for example
 with  contact interaction, Lattice QCD and quark determinant
frameworks.
In the CQM the axial pion-constituent quark 
coupling 
 constant  is usually taken to be $g_A=1$
\cite{Weinberg2010} whereas the  
 pseudoscalar coupling constant ($g_{ps}$) one
may be considered in two different ways.
Either the pion-nucleon coupling constant is divided by 
the number of constituent quarks, i.e. $g_{ps}/3$, or 
as it is considered as an average of the pion coupling to each 
of the quarks, with a
 a phenomenological value in the range $13$-$13.5$ \cite{Gps}.
Among the most useful low energy 
theorems in hadron dynamics,   
the Goldberger-Treiman 
 relation that relates 
the  
 chiral limit  pion nucleon pseudoscalar coupling constant 
to the axial one \cite{GTR,gernot-fischer}:
$g_{ps} f_\pi = g_A M_{nuc}$, in terms of the nucleon mass
($M_{nuc}$) and the pion decay constant ($f_\pi$).
The behavior of pion properties
became of high interest with the expectation 
of understanding hadron dynamics in the presence of 
magnetic fields.
 Should a magnetic field induce different pion couplings to the constituent quarks,
the corresponding pion internal structure 
is modified.

Magnetic field 
corrections to usual couplings already present in the vacuum,
such as the pion axial and pseudoscalar
 couplings to nucleons or to constituent quarks,
have been calculated
recently within different approaches 
and, usually, they   lead to considerably spatial  anisotropies 
although isotropic components also show up
 \cite{PPNP,couplings,Braghin2020,yukawaB2024,Villavicencio,Dominguez:2023bjb}.
Theoretical estimates and 
assessments based on experimental  data
indicate that, although initial magnetic field, rather  in the quark gluon  phase,
can be very strong,
it decays very fast so that in the hadron phase
it should not be really too strong
\cite{PPNP,Zwang-etal-B}.
However,
magnetic fields have also been found to be associated to further
 unusual couplings, which would violate conservation rules in the vacuum,
 for example, in \cite{PPNP,carlomagno-etal}.
In spite of the difficulty in verifying experimentally
possible effects of these couplings, one may
expect that 
 specific observables could, possibly, allow the identification 
of their effects as signature of magnetic fields in r.h.i.c.
With progressive improvement  of experimental
precision, one could possibly detect
   hadrons,
  formed  in  magnetic fields that,
would be in excited states when leaving  the magnetic field region, 
and subsequently decay from such excited state, or decay
in the magnetic field, what could be 
observed.
Effects of these types might possibly lead to signatures of 
the presence of magnetic field in the confined (hadronized)
 phase
of  r.h.i.c.

In this work, 
 (single-)
pion couplings to vector and scalar  constituent  
quark currents induced by an external background magnetic field,
relatively weak with respect to the quark (constituent) mass,
will be calculated.
The resulting  form factors  will be found to be 
(linearly)
proportional to magnetic field and represent the most important results.
 From the resulting equations, we propose the resulting interactions correspond to the following Lagrangian terms
\begin{equation}
    {\cal L}_1\sim 
    \frac{g_1}{\bar{M}^3}
    \tilde F^{\mu\nu}\partial_\mu\pi^i\,\bar q \tau^i \gamma_\nu q,
    \quad 
    {\cal L}_2\sim
    \frac{g_2}{\bar{M}^4} \tilde F^{\mu\nu}\partial_\mu\pi^i\,\bar q \tau^i\partial_\nu  q
\end{equation}
 where $\tilde{F}^{\mu\nu}$ is the dual electromagnetic stress tensor, $\bar{M}$ is some mass scale and  $\tau_i$
 are the isospin Pauli matrices.
We will employ a diagrammatic method
to compute   weak magnetic field induced
pion-constituent quark form factors 
that vanish in the vacuum.
The usual pseudoscalar pion-constituent quarks coupling will be employed.
This coupling can be considered 
as arising from  Weinberg's Large Nc Effective Field Theory (EFT) \cite{Weinberg2010}
provided a redefinition of the pion field leading to linear realization of  chiral symmetry.
This is  seen in the derivation given in Refs.
\cite{BraghinEFT} by starting with a quark-antiquark interaction mediated by gluon exchange.
This EFT is defined in terms of constituent quarks and gluons and the pion field in the non-linear realization. 
Therefore, it presents all the needed ingredients for the calculations we consider.
The pseudoscalar pion coupling to quarks (or nucleon) appears 
phenomenologically
in the quark- Linear Sigma-type models (LSMq)
\cite{LSM}.
%In the present work, quark and gluon interaction and the
%(effective) gluon propagator
%are input  for the calculation with QCD-type Feynman rules.
 The three-point Green's function
will be projected into the Dirac-flavor channels whose
counterparts in the vacuum are zero, 
namely  scalar and  vector single pion couplings to constituent quarks.
Accordingly,
 semi-analytical equations for the corresponding form factor
will be obtained, and numerical results will be  shown for two kinematical regions:
on-shell constituent quarks at rest with off-shell pion,
and off-shell constituent quark with on-shell massless pion.
The neutral pion is associated to the isotriplet state P$_3$. Results are also valid for the $\eta$ meson if considered as primarily described by 
U(3) flavor state P$_8$.
The results might be valid for $\eta'$ if meson mixing is considered for the flavor state 
P$_0$.
In the next section, the corresponding three-leg
Feynman diagrams are   calculated and projected in the 
two flavor-Dirac channels of interest.
In the following section, numerical results are  presented,
for each of the kinematical regions, and discussed.
Some possible phenomenological implications are   discussed.
In the last section, a summary and discussion are   presented.

\section{ Pion-quark interaction channels}

\begin{figure}[ht!]
\centering
\includegraphics[width=50mm]{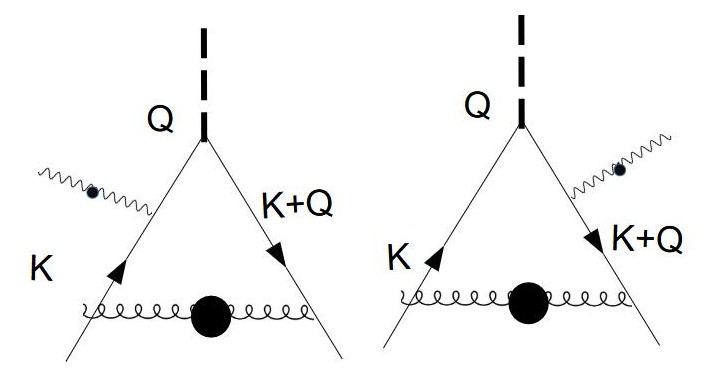}
 \caption{ 
\small
Constituent quark- pion interaction in a magnetic field:
dashed line for pion,  solid line for quark,
 wiggly lines with a  dot for gluons, soft wiggly line for the magnetic field insertion.
}
\label{fig:pion-CQ-B}
\end{figure}

In this section the two diagrams of Fig. \eqref{fig:pion-CQ-B}
are calculated and projected into 
scalar and vector channels,
 corresponding to 
magnetic field induced couplings since they disappear in the 
limit of $B=0$.
The pseudoscalar pion coupling to (constituent) quarks 
will be considered. 
The theoretical framework we will consider for the calculation 
of the anomalous pion couplings  relies on the 
Large Nc Weinberg Effective Field Theory (EFT) \cite{Weinberg2010}
that, in a particular dimensionless 
definition of pion field, written in terms of the 
covariant derivatives \cite{Weinberg-book},  and with the electromagnetic coupling,
can be written as:
\begin{eqnarray}
{\cal L}_{eff}^{(2)} &=& 
\bar{\psi} \left(  i \gamma \cdot {\cal D}
- M \right) \psi
- 2 g_A  ( 
D_\mu {\pi}_i) 
   \bar\psi   i \gamma_5 \gamma^\mu \tau_i \psi
- \frac{1}{4}Tr {\cal G^{\mu\nu} } {\cal G_{\mu\nu}}
+ 
{\cal L}_{ChPT}
\end{eqnarray}
where ${\cal G^{\mu\nu}}$ is the gluon field strength
for the gluon field ${\cal A}^\mu$, $g_A$ is 
the axial pion coupling to quarks, 
the photon field 
is $A^\mu$, ${\cal L}_{ChPT}$
encodes all the terms from Chiral Perturbation Theory that will
not be used, and 
the following covariant derivative that includes  the pion, gluon and 
photon fields is defined as:
$
{\cal D}^\mu \psi  = \left[ \partial^\mu  
 - i e A^\mu - i g {\cal A}^\mu 
+ 2 i   g_v
D_\mu \vec{\pi}
\cdot (\bar\psi   \gamma^\mu  \vec{\tau}  )
\right] \psi$,
where the usual pion vector coupling to quarks appear with the pion covariant derivative $D_\mu \vec{\pi} = \frac{\vec{\pi} \times \partial_\mu \vec{\pi}}{1+{\vec{\pi}}^2} $.
This Lagrangian is written in terms of a non-linear definition of the 
pion field which can be redefined such that the pseudoscalar pion coupling
to quarks emerges. 
This is somewhat exploited in a derivation of this 
EFT provided in  \cite{BraghinEFT}.
As a consequence, instead of the 
vector and axial pion-quark couplings, the following one will be considered:
\begin{eqnarray}
{\cal L}_{\pi-q} = g_{ps} \;
\bar{\psi} %% ( i \slashed{D} + 
i \vec{\pi} \cdot \vec{\tau}  \psi , 
\end{eqnarray}
where  $g_{ps}$ is the pion pseudoscalar coupling to (constituent) quarks.
The gluon sector   gives rise to an effective gluon propagator that can 
be parameterized by a 
gluon effective mass.
The gluon propagator will not be calculated in this work, it will be considered as an external input.
To associate the involved quarks to constituent quarks
the gluon exchange will be included 
in the pion-quark form factor and, to take into account non-Abelian effects, 
an effective  gluon propagator  will be considered. 
%This approach is similar  to the calculation of
%form factors  from  a quark
% determinant in the presence of local meson fields and 
%(constituent) quark currents \cite{Braghin2018b,Braghin2020}.
The form factors that will be calculated are depicted in  Fig. \eqref{fig:pion-CQ-B} where the internal quark lines (solid lines) 
are the only receiving background
magnetic field insertion - the wavy line with a dot. 
The leading contribution for this effect is linear in the magnetic 
field as shown below.
 The gluon lines with a full dot 
correspond to an effective gluon propagator whose magnetic field
corrections would require to consider a quark-antiquark polarization process
which would make this contribution   sub-leading.
The external pion line is the dashed line.
Weak magnetic field 
for the pion pseudoscalar coupling to constituent quarks have been computed 
by considering the quark determinant in the presence of 
quark currents and local meson fields.
 The leading contribution
was found to be quadratic in the magnetic field \cite{Braghin2020} 
which therefore is non-leading, and it will not be considered.

\FloatBarrier

Form factors for the diagrams above, computed 
with the pion pseudoscalar coupling 
and obtained by a projection of the vertex of the figure above, respectively in the scalar and vector channels \cite{Itzykson-Zuber}, can be written as::
\begin{eqnarray} \label{PiSps}
\Pi_{S,ps}^i 
&=& G_{ps}
 Tr \; \left[ {\cal P}_s^\pi \;
\Gamma_{\mu,a} S (k+K)  \; ( i \gamma_5 \tau_i ) \;
S (k+K+Q)
\Gamma_{\nu,a} R^{\mu\nu}_{ab} (-k)
\right],
\\ \label{PiVecps}
\Pi_{V,ps}^{i,\rho }
&=& G_{ps} Tr \; \left[ 
{\cal P}_v^{\rho,\pi}
 \;\Gamma_{\mu,a} S (k+K) \; (  i \gamma_5 \tau_i  ) \;
S (k+K+Q)
\Gamma_{\nu,a} R^{\mu\nu}_{ab} (-k)
\right]
\end{eqnarray}
where $G_{ps}$ corresponds to the pion pseudoscalar coupling coupling constant, and where $\tau$ are the Pauli matrices,
being that
 neutral pion involves $i=3$ and charged pion combinations of $i=1$ and $i=2$.
 The
trace $Tr$ is a generalized trace in Dirac indeces, flavor and color indices as well as an integration in internal momentum $k$.
The indexes $a,b=1,...8$ are the color indices,
and where the quark-gluon vertex is 
%\begin{eqnarray}
$\Gamma^\alpha_a = - i g 
\frac{\gamma^\alpha \lambda_a}{2}$,
%\end{eqnarray}
and
$ R^{\mu\nu}_{ab} (-k)$ is a gluon propagator written below,  that will incorporate the multiplicative factor $g^2$ from the quark-gluon vertices.
 The operators ${\cal P}^\pi$ project into the scalar and vector channels
and in the pion charge eigenstate are given by:
\begin{eqnarray}
{\cal P}_s^\pi &=& \frac{1}{4} \; 1_{4} \;\tau_\pi,
\;\;\;\;\;\;
\left({\cal P}_v^{\pi}\right)^\rho =
\frac{1}{4}
\gamma^\rho \tau_\pi,
\end{eqnarray}
where the isospin projectors for neutral and charged pion 
given in $\tau_{\pi^\pm} = \frac{1}{\sqrt{2}}(\tau_1\pm i\tau_2)$ in the case of charged pions and $\tau_{\pi^0}=\tau_3$ for neutral pions.
The traces in Eq\,\eqref{PiSps}  and \eqref{PiVecps} with the  ${\cal P}^\pi$ operators vanish in vacuum.
This way, the other linear independent structures of the vertex are eliminated.
In other words, the trace over the projector times the corrected vertex isolates components in Clifford basis.
The magnetic field insertion provides non-vanishing contributions for vector and scalar components.

By considering flavor U(3)
pseudoscalar meson couplings to constituent quarks, the same development described here is also valid for the $\eta$ meson if described by the flavor state P$_8$.
The $\eta'$ meson is usually described by the state P$_0$
which, however, will only provide non-zero couplings of the type discussed in this work if meson mixing to $\eta$ and $\pi^0$ are also considered.
This will be discussed in another work.

In these diagrams, the quark propagator
is immersed in a magnetic field.
Interestingly, for the type of diagram shown in Fig. \eqref{fig:pion-CQ-B},
the Schwinger phase can be gauged away, what simplifies the
calculation \cite{SchwingerPhase}.
 For the limit of weak 
magnetic field  with respect to the (constituent)  quark mass, 
it can be written as \cite{weakB1}
\begin{eqnarray}  \label{quark-prop}
S (k) &=&  S_0 (k) + S_1 (k) (e B_0) = i 
\frac{ \slashed{k} + {M} }{ k^2 - {M}^2 } 
- \gamma_1 \gamma_2
 \frac{
 {M}^2(  \gamma_0 k^0 - \gamma_3 k^3  +  {M}  )
 }{
(k^2 - {M}^2)^2 }  \hat{Q} \frac{e B_0
}{ {M}^2}
\end{eqnarray}
where 
$\hat{Q}$ is the quark charge matrix and
$\tilde{S}_0(k) = 1/(k^2 - M^2)$

 The effective (confining)  gluon propagator 
employed was inspired in \cite{cornwall}
and is given by:
\begin{eqnarray}
\label{gluonprop}
R^{\mu\nu}_{ab} (k) \equiv i 
\delta_{ab} g^{\mu\nu}
K_F
R(k) =  i 
\delta_{ab} g^{\mu\nu}
K_F \frac{ 1}{ ( k^2  - M_g^2)^2 },
\end{eqnarray}
where $K_F = 8 \pi^2 M^2$ is the corresponding
 string tension
and $M_g$ is the effective gluon mass.
By taking into account both contributions  of the magnetic field insertions from both diagrams \eqref{fig:pion-CQ-B},
it yields, for the charged pion case,
\begin{align}   \label{Pisp}     
    i \Pi_\text{\tiny S(P)}^i &= i
    G_{ps}\frac{4
      \; (N_c^2-1) \, p\cdot\tilde Q_\parallel }{4 [p^2-M^2][(Q+p)^2-M^2]}\left[\frac{e_u B}{p^2-M^2}+\frac{e_d B}{(Q+p)^2-M^2}\right]
    R(-p-K) 
      \\   
     i \Pi^{\mu,i}_\text{\tiny V(P)} &=  i G_{ps}\frac{2 (N_c^2-1)  \,M \tilde Q_\parallel^\mu }{4 [p^2-M^2][(Q+p)^2-M^2]}\left[\frac{e_u B}{p^2-M^2}+\frac{e_d B}{(Q+p)^2-M^2}\right]
       R(-p-K)
\end{align}
with $\tilde Q_\parallel^\mu = \epsilon^{12\mu \nu} Q_\nu$ and  $\tilde p_\parallel^\mu = \epsilon^{12\mu \nu} p_\nu$ and where the quark charges are
$e_u= \frac{2 e}{3}$ and  $e_d=- \frac{e}{3}$ being $e$ the  proton charge.
 For the neutral pion, is the same result, but replacing  $e_u\to \frac{e}{\sqrt{2}}$ and  $e_d\to \frac{e}{\sqrt{2}}$.

By employing the Feynman parameterization 
to perform the momentum integrals,
these form factors can be written in terms of  
integrations of two Feynman parameters, and they read
\begin{eqnarray} \label{Fsppi}
\Pi_{S(P)}^{\pi} \equiv 
F_{S(P)}^\pi (Q,K) &=&
\frac{  4 G_{ps} (N_c^2 - 1) M^2 K_F }{ 4   (4\pi)^2 } \frac{eB}{M^2}   
\; I_S^\pi 
\\ 
\label{Fvppi}
\Pi_{V(P)}^{\mu,\pi}
\equiv 
 \left[   \tilde{Q}^\mu \right]  
  F_{V(P)}^\pi (K,Q) &=& 
\frac{ 2 G_{ps} (N_c^2 - 1) M^3 K_F }{ 4 (4\pi)^2 } \frac{eB}{M^2} \;  \tilde{Q}^\mu    \; 
\; I_V^\pi  
,
\end{eqnarray}
%where the sufix $^\pi$ stands for $\pi = 0,c$ the neutral and charged pion, 
with the momentum-dependent integrals given by:
\begin{eqnarray}
I_{S}^\pi 
&=&
2 \int_0^1 d z \int_{0}^{1-z} dy
\; t
\left[
%\frac{  
%\tilde{Q}_\parallel \cdot [ - V (e_u + e_d) ] }{
%D^3}
- \frac{3 V \cdot \tilde{Q}_\parallel (e_u + e_d) }{
D^3}
+ 
\frac{3  V \cdot \tilde{Q}_\parallel [ e_u (Q^2 - 2 V \cdot Q_\parallel )
+ (V^2 - M^2 ) (e_u + e_d) ]}{D^4}
\right]
\nonumber
\\
I_V^\pi 
&=& -  \int_0^1 d z \int_{0}^{1-z} dy  \;\;t
\left[ - 
\frac{4 (e_u + e_d)}{ D^3} +
\frac{ 6 \left(
(e_u+e_d)[ - M^2 + V^2 ]
+ e_u [ Q^2 - 2 Q \cdot V ]
\right)}{D^4}
\right]
\end{eqnarray}
where the following quantities have been defined
\begin{eqnarray}
V^\rho &=& y Q^\rho - z K^\rho,
\\
D &=& [ Q^2 y(y-1) + K^2 z (z-1) 
- 2 K \cdot Q y z + M^2 (1-z) + M_g^2 z]
\\
t &=& y z (1-y-z).
\end{eqnarray}
It is seen in Eqs. \eqref{Fsppi} and \eqref{Fvppi} that these form factors 
depend linearly in $(eB)/M^2$.

  The scalar form factor has also been
 calculated by considering the axial
pion coupling to quarks, i.e.
$\gamma_\rho \gamma_5 \frac{Q^\rho}{f_\pi}  \tau_i$,
 and it is given by:
\begin{eqnarray}
\label{Pi-S-A}
i \Pi_{S,(A)}^i &=& g_A Tr \; \left[ 
{\cal P}_s^i
 \;
\Gamma_{\mu,a} S  (k+K) 
\frac{i Q^\rho}{f_{\pi}}\gamma_\rho \gamma_5 \tau_i S  (k+K+Q) 
\Gamma_{\nu,a} R^{\mu\nu}_{ab} (-k)
\right] ,
\end{eqnarray}

This form factor can be simply related to the 
scalar one calculated with the pseudoscalar pion-quark coupling by
%\begin{eqnarray}
%\label{GTR}
$\Pi_{S(A)}^{\pi} = \frac{ 2 g_A M}{ G_{ps} f_\pi }
\Pi_{S(P)}^\pi$.
%\end{eqnarray}
This coefficient in the r.h.s. is 2/3
if we consider the GT-relation 
$(3 g_A M /G_{ps} f_\pi \simeq 1)$ 
\cite{GTR}
at the quark level.

\section{ Numerical results}
\label{sec:numerics}

For numerical calculations the following values 
have been adopted for the constituent quark mass
and coupling constants:
\begin{eqnarray}
M_g = 400 MeV && M  = 350 MeV,
\;\;\;\;\; G_{ps} = 13,
\end{eqnarray}
where
the value for   $G_{ps}$ is usually considered in phenomenological investigations of the pion-nucleon interaction \cite{Gps}.
We will explore two kinematical regions, which are 
\begin{itemize}
\item On-shell quarks and off-shell pions
%\begin{eqnarray}
%\label{kinematical-1}
$K^\mu = ( M , 0,0,0 )$, 
and $Q^\mu = (0,-\vec{Q})$,
with
$Q^2 = - \vec{q}^{\; 2}$.
%\end{eqnarray}
This can be interesting for analyzing processes such as pion exchange
in which constituent quarks are representative of the nucleon
(or any other hadron) - along with the usual view of the
CQM.

\item Off-shell quarks and massless on-shell pions
%\begin{eqnarray}
%\label{kinematical-2}
for $K^\mu = ( 0, - \vec{K} )$ and  $Q^\mu = ( |\vec{Q}|,-\vec{Q})$.
%\end{eqnarray}
With this, one may somewhat expect to provide a model 
for a pion  absorbed by or emitted  from a 
confined, or at least not properly propagating, 
constituent quark.

In this case, some values
for the incoming quark momentum
  have been considered:
%\begin{eqnarray}
%\label{constraints}
$K_3 = \frac{Q_3}{2}$,
%\;\;\;\;\;
$|\vec{K}| = \frac{|\vec{Q}|}{2}$,
and
 $\vec{K} \cdot \vec{Q} = 0$.
%\end{eqnarray}
\end{itemize}
In all cases only transversal pion momentum
with respect to  incoming quark momentum will be considered.
More detailed analysis of the results will be provided elsewhere.

\subsection{ On-shell constituent quark and off-shell pion}

\begin{figure}[H]
\centering
\begin{minipage}{.4\textwidth}
  \centering \includegraphics[width=.9\linewidth]{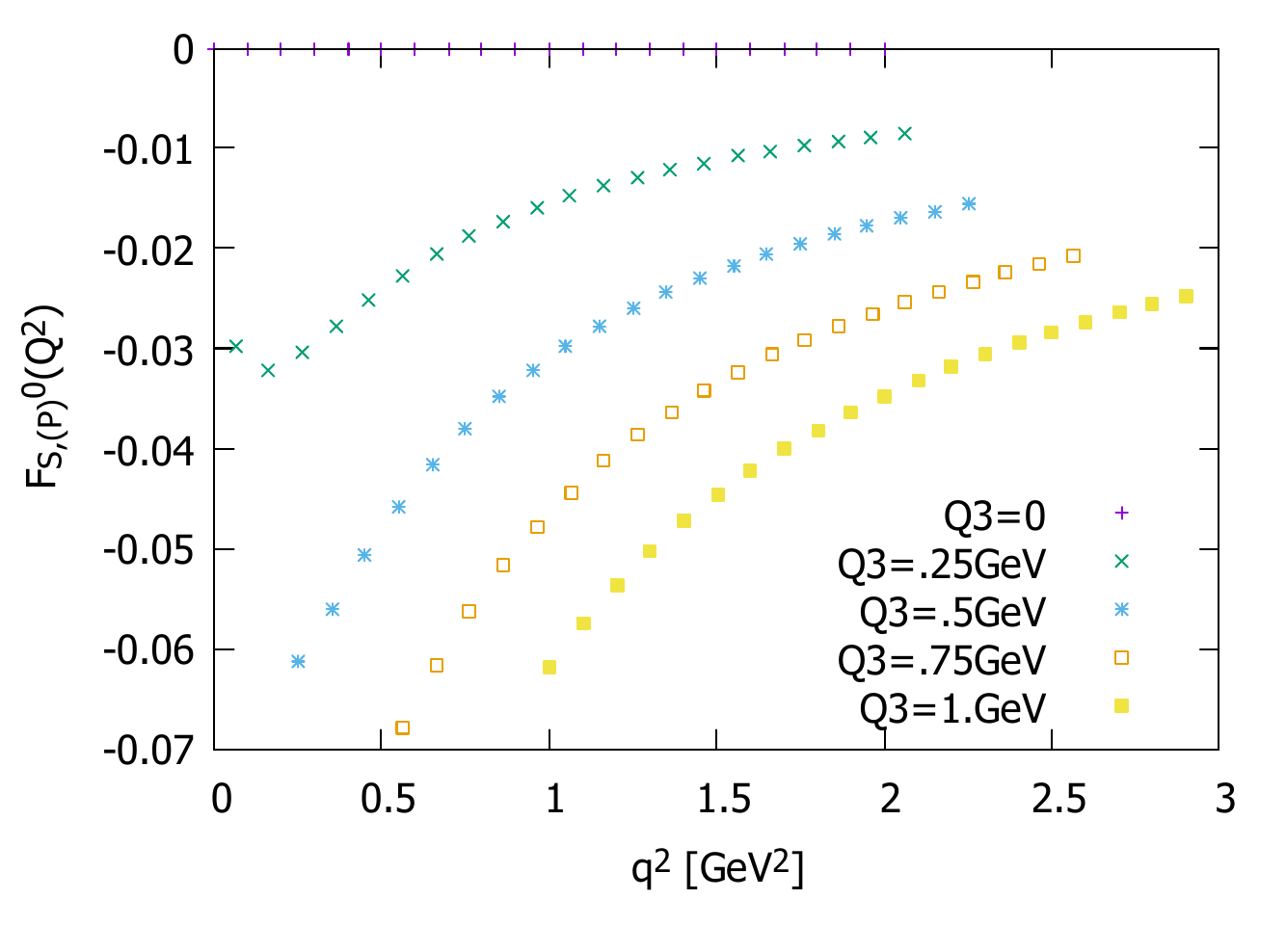}
  \caption{  Neutral pion scalar form factor, Eq. \eqref{Fsppi}, for on-shell quark and off-shell pion.}
  \label{fig:Sps-neu}
\end{minipage}%
\hspace{1cm}
\begin{minipage}{.4\textwidth}
  \centering
  \includegraphics[width=.9\linewidth]{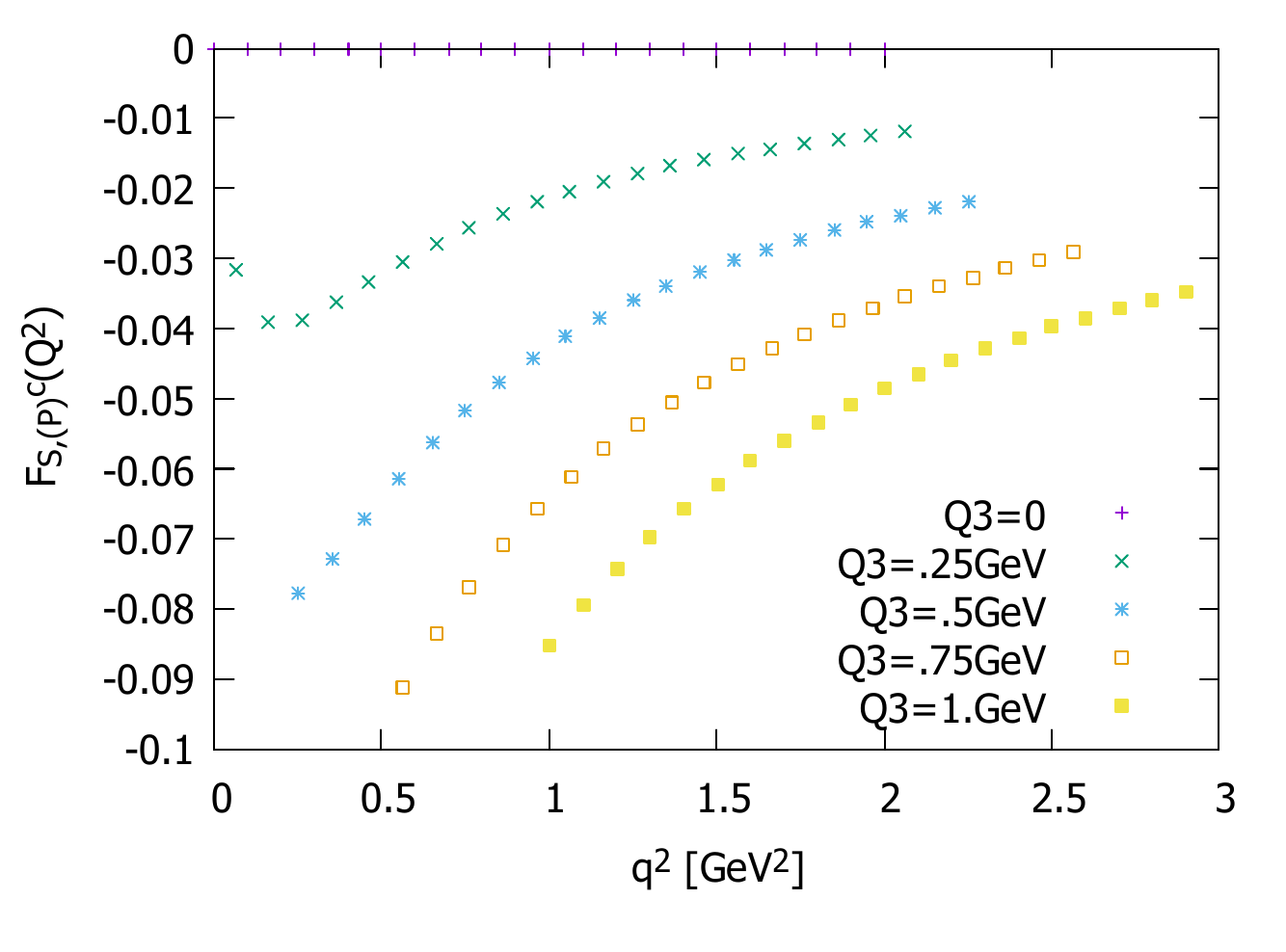}
  \caption{Charged
  pion scalar form factor, Eq. \eqref{Fsppi}, for on-shell quark and off-shell pion.}
  \label{fig:Sps-cha}
\end{minipage}
\end{figure}

This kinematical region might be typical for pion exchange potential calculations.
In Figs. \eqref{fig:Sps-neu} and \eqref{fig:Sps-cha} the
scalar form factor, for  on-shell constituent quark and off-shell pion,
 is displayed respectively for the neutral and charged pions.
Different values of the longitudinal pion 3-momentum $Q_3$ up to 1 GeV are considered.
This form factor is proportional to $Q_3$ so that for a purely transversal pion momentum 
the form factor is zero.
The overall 
strength  is larger for the charged pion and the decreasing behavior with $q^2=-Q^2$ is similar for both cases. 
For increasing values of $Q_3$, 
the larger values of the form factor
 might be bounded to some value because it 
increases less for larger $Q_3$.
We cannot expect this one-loop form factor  to be 
interesting for larger momenta since higher order contributions
should be relevant.

\begin{figure}[H]
\centering
\begin{minipage}{.4\textwidth}
  \centering 
\includegraphics[width=.9\linewidth]{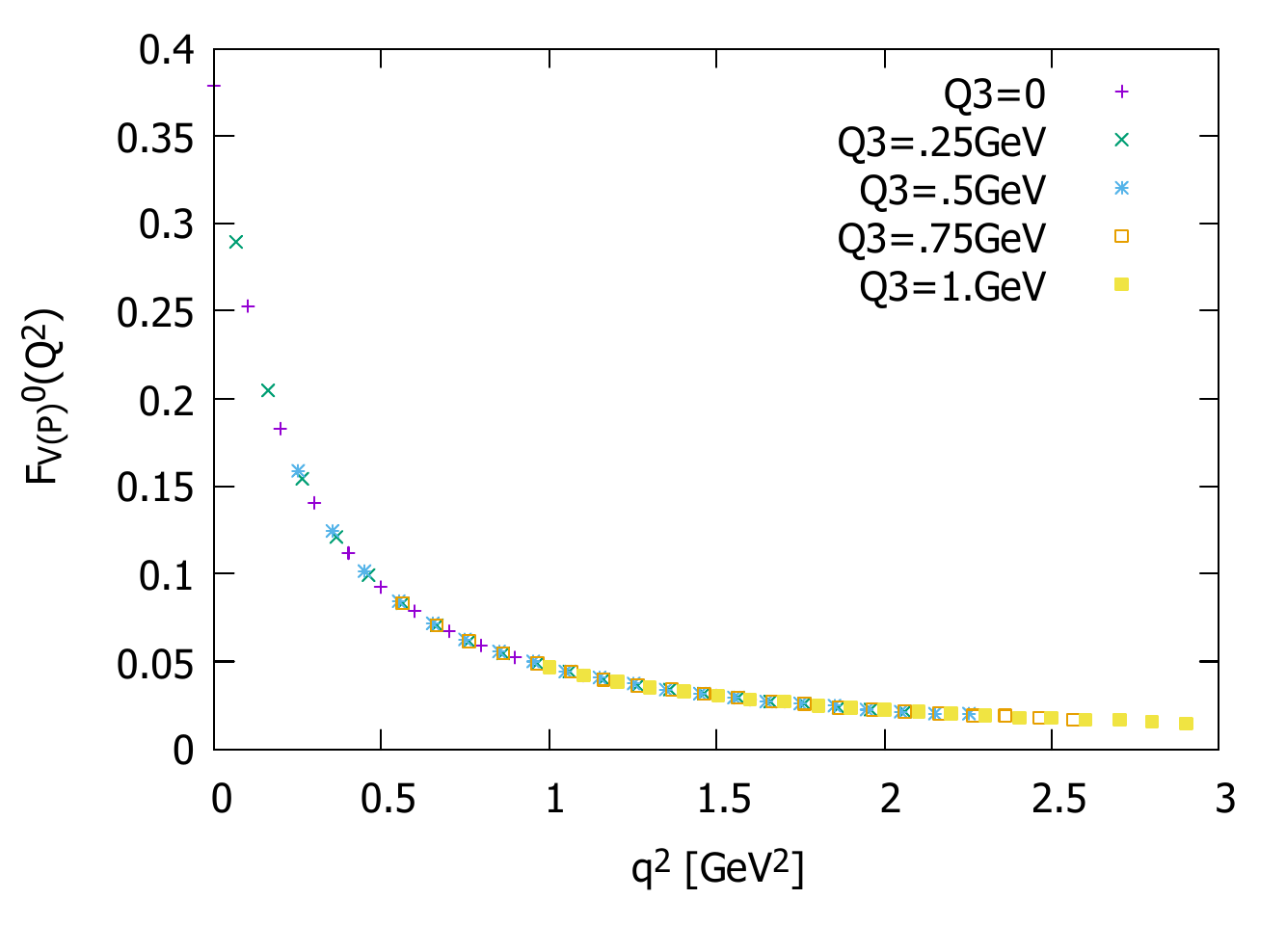}
  \caption{  Neutral pion vector form factor, Eq. \eqref{Fvppi}, for on-shell quark and off-shell pion.}
  \label{fig:Vps-neu-mu0}
\end{minipage}%
\hspace{1cm}
\begin{minipage}{.4\textwidth}
  \centering
  \includegraphics[width=.9\linewidth]{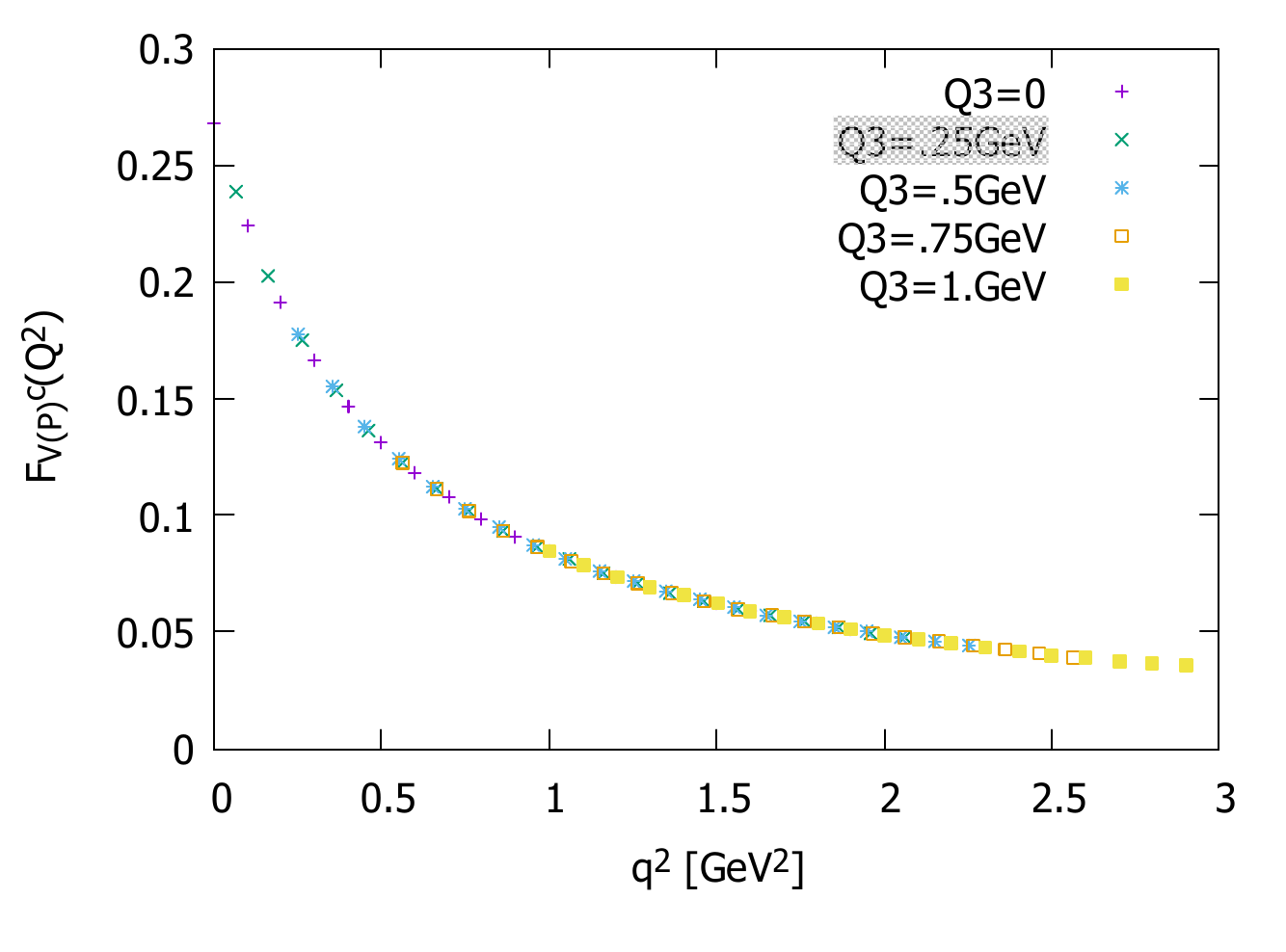}
  \caption{ Charged 
  pion vector form factor, Eq. \eqref{Fvppi}, for on-shell quark and off-shell pion.}
  \label{fig:Vps-cha-mu0}
\end{minipage}
\end{figure}

The vector form factor, by means of $F_{V(P)}(Q^2)$
given in eq. \eqref{Fvppi},  without the asymmetric 
factor
$\tilde{Q}^\mu$,
is shown in 
Figures \eqref{fig:Vps-neu-mu0} and \eqref{fig:Vps-cha-mu0}, respectively, for neutral and charged pions.
This quantity, $F_{V(P)}$, is therefore 
fully isotropic.
In contrast to the scalar form factor, the neutral pion vector form factor has a larger strength than the charged pion.
The overall decrease for increasing $q^2$ is faster for the neutral pion, with smaller values at larger $q^2$,
 and this indicates a slightly shorter range behavior.
The vector form factor  in the figures is not multiplied by $\tilde{Q}^\mu$, 
being this component  larger in strength than the scalar one
at smaller $q^2$.
The diverse positive and negative signs of scalar and vector channels
suggest it is  possible to generate a (very weak) component of repulsive 
constituent quark interaction for pion exchange.

\subsection{ Off-shell constituent quark and on-shell
massless pion}

For on-shell  pion, the corresponding processes
of diagrams \eqref{fig:pion-CQ-B} can represent pion emission or absorption.
by constituent quarks.
Pion decay into quark-antiquark scalar or vector currents, however, would  not be allowed due to energy conservation.
The analysis for this kinematical region is therefore guided
by these processes.
For this, the massless pion case with off-shell quarks
will be considered.
Besides that
 pion and constituent quark three-momenta will be
considered  transversal as mentioned.
As a result there are subtle differences between 
pion emission and absorption
with the change of $\vec{Q} \to - \vec{Q}$ in 
Eqs. \eqref{Fsppi} and 
\eqref{Fvppi}.

\begin{figure}[H]
\centering
\begin{minipage}{.4\textwidth}
  \centering \includegraphics[width=.9\linewidth]{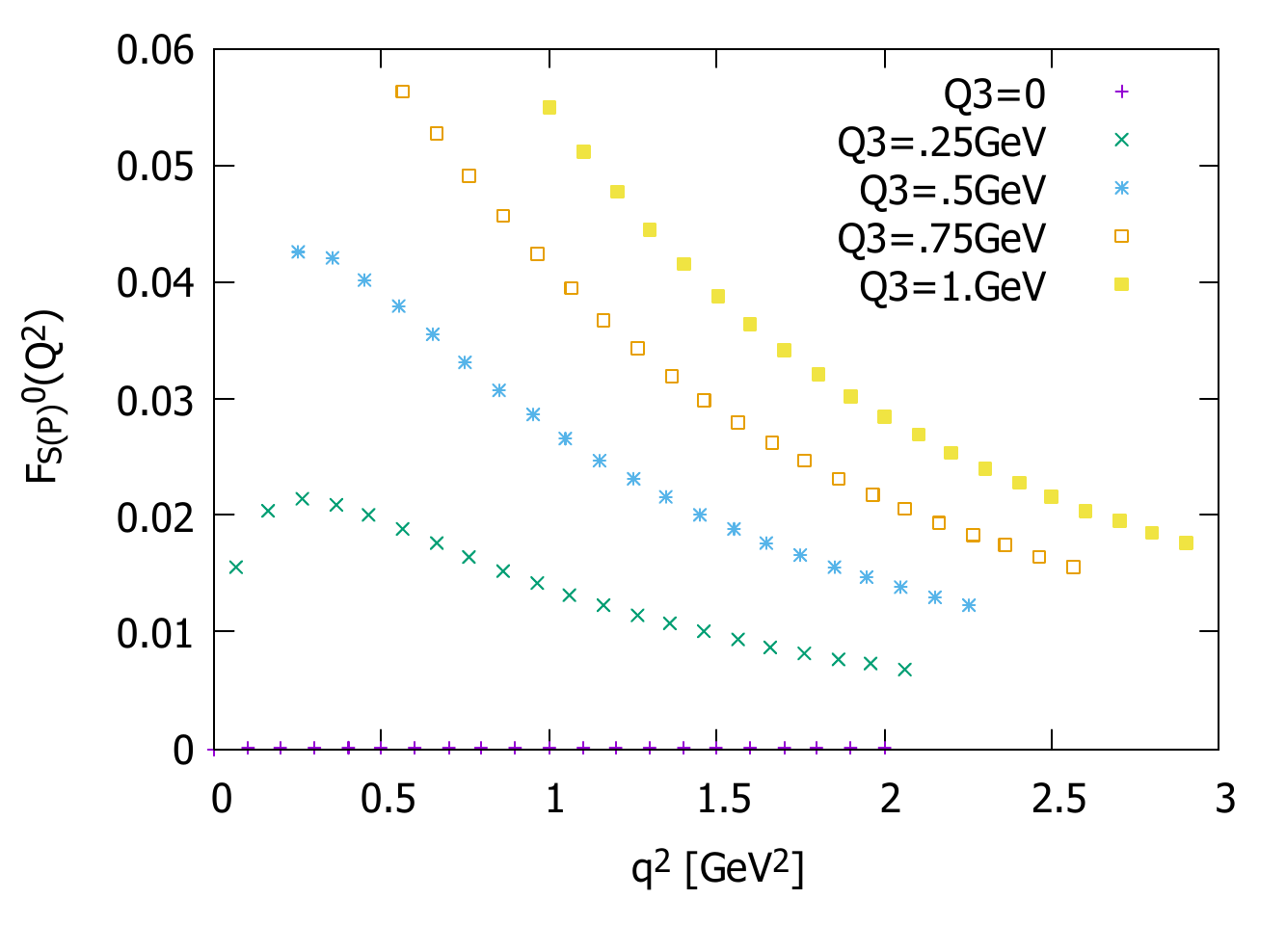}
  \caption{
   Neutral pion scalar form factor, Eq. \eqref{Fsppi}, for off  shell quark and massless on-shell pion.}
  \label{fig:Sps-qofpon-neu}
\end{minipage}%
\hspace{1cm}
\begin{minipage}{.4\textwidth}
  \centering
  \includegraphics[width=.9\linewidth]{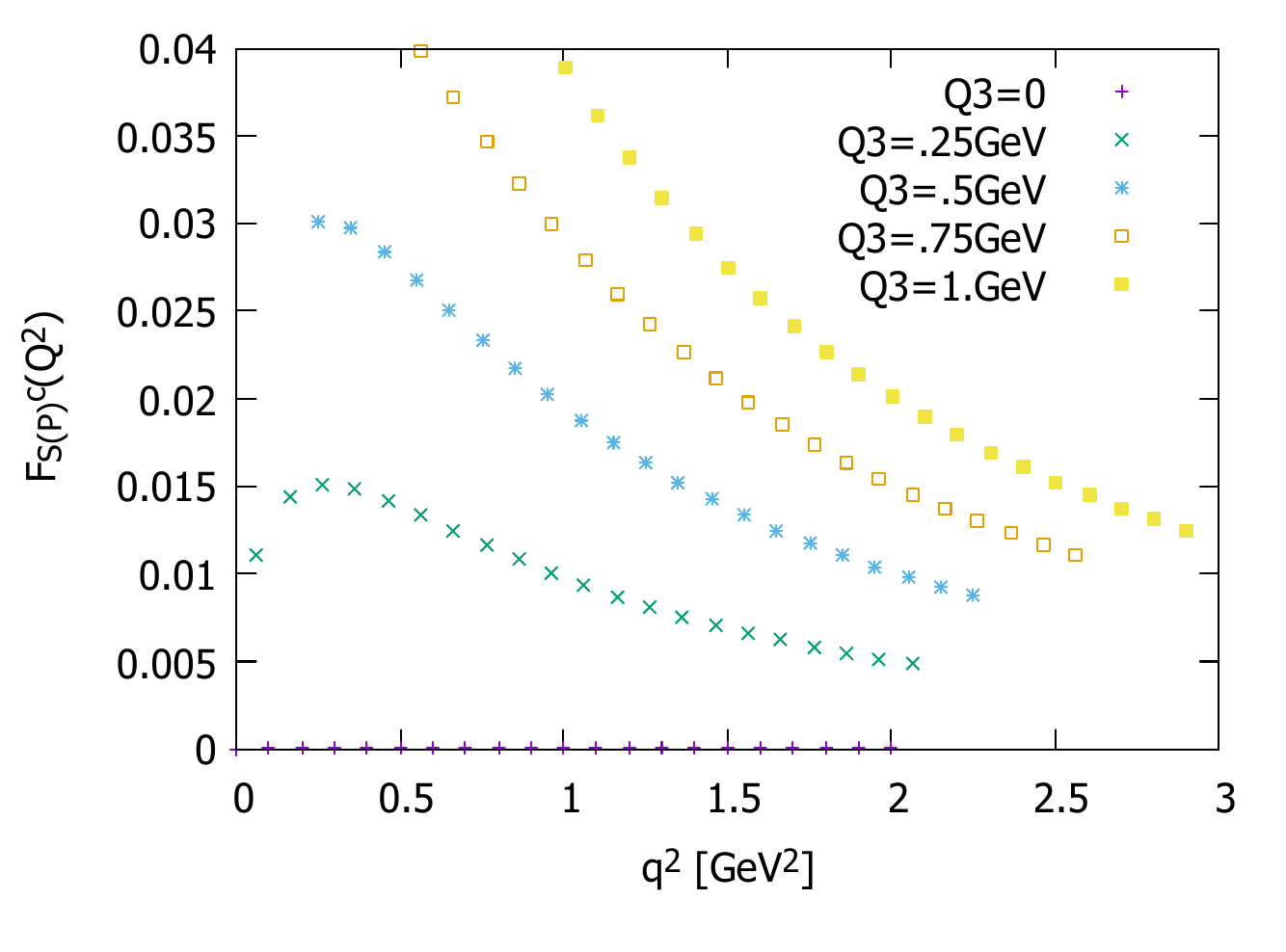}
  \caption{Charged pion scalar form factor, Eq. \eqref{Fsppi}, for off  shell quark and massless on-shell pion}
  \label{fig:Sps-qofpon-cha}
\end{minipage}
\end{figure}

In Figs. \eqref{fig:Sps-qofpon-neu}
and \eqref{fig:Sps-qofpon-cha}
the scalar form factors, respectively
 for neutral and charged pions, 
are displayed for massless on-shell pion absorbed or 
emitted by an off-shell constituent quark, 
as  discussed above.
The strength of the neutral pion form factor is 
larger than the charged pion one - contrary to the 
off-shell pion discussed above.
Besides that, for larger $q^2$ the neutral pion form factor
reaches smaller values leading to a longer range.
The sign of the form factor turns out to be negative.
Neutral and charged pion 
have both form factors
 proportional to $\tilde{Q}$
and have the trend to increase for larger $Q_3$,
although they decrease for larger  $q^2$.
Also, the increase with larger $Q_3$ is progressively 
weaker for larger $Q_3$ and, possibly,  it suggests it 
might reach a maximum
strength for larger $Q_3$.

\begin{figure}[H]
\centering
\begin{minipage}{.4\textwidth}
  \centering \includegraphics[width=.9\linewidth]{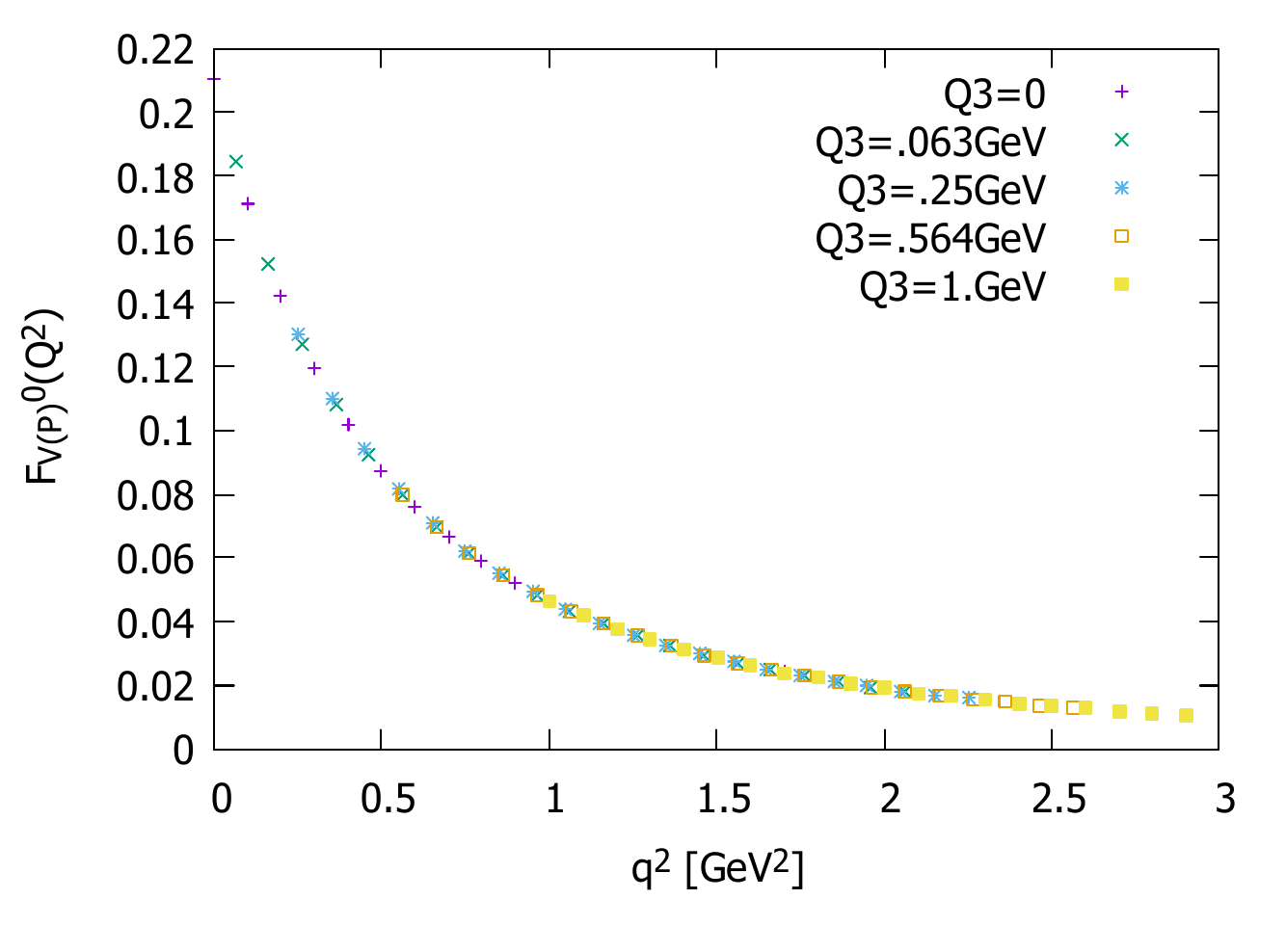}
 \caption{ 
Neutral pion vector form factor, Eq. \eqref{Fvppi}, for off  shell quark and massless on-shell pion,}
  \label{fig:Vps-qof-neu-semmu}
\end{minipage}%
\hspace{1cm}
\begin{minipage}{.4\textwidth}
  \centering
  \includegraphics[width=.9\linewidth]{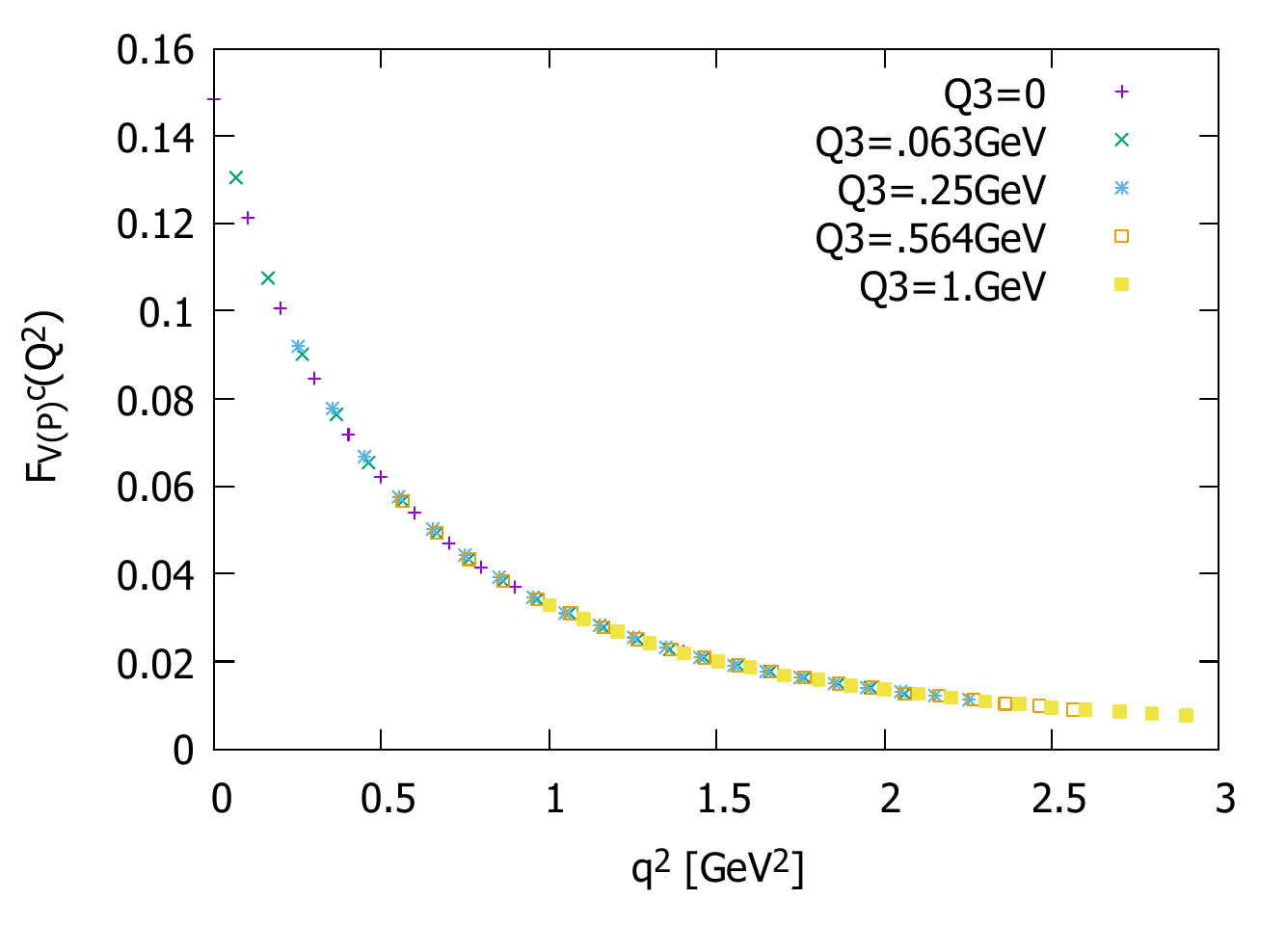}
 \caption{ 
 Charged  pion vector form factor, Eq. \eqref{Fvppi}, for off  shell quark and massless on-shell pion. }
  \label{fig:Vps-qof-cha-semmu}
\end{minipage}
\end{figure}

In Figs. \eqref{fig:Vps-qof-neu-semmu}
and \eqref{fig:Vps-qof-cha-semmu}
the vector  form factors $F_{V(P)}(Q^2)$, respectively
 for neutral and charged pion, 
for the same case of the previous figures,
are displayed. 
Again, the form factors are  shown  without
the common multiplicative factor $\tilde{Q}^\mu$ and 
the dependence on the momentum asymmetry,
i.e. on $Q_3$ or $P_3$, does not show up.
Due to this reason, the vector neutral pion form factor is somewhat 
larger than the charged one for 
small $q^2$.
However, the neutral pion form factor
 rapidly decay  to smaller values 
and, 
by $q^2 \sim 2$ GeV$^{2}$,
reaches nearly the same values as the charged pion one.
The vector neutral or charged pion form factors go to zero faster than the scalar one,
indicating shorter range interaction, although this is
 dependent
on the momentum asymmetry by $|Q_3|$.

% \subsection{ Average quadratic radia 
%(a.q.r.)}

For the scalar and form factors, the following forms were used to make possible the
definition of the a.q.r. - by factorizing the dependence on the 
quark and pion momenta
as displayed in Eqs. \eqref{Pisp} and \eqref{Fsppi}:
$\Pi_{S(P)}^\pi \equiv  G_{S(P)}^\pi  V \cdot \tilde{Q}_\parallel$,
and 
$
\Pi_{V(P)}^{\mu,\pi} \equiv  G_{V(P)}^\pi  M \tilde{Q}^\mu$.
However, for the sake of extracting numerical results for the a.q.r., these momenta will be taken to be $|Q| = |K| = M$:
\begin{eqnarray}
< r^2 >_{S(P)}^\pi &=& - 6 \left. \frac{ d \Pi_{S(P)}^\pi }{d Q^2} \right|_{Q^2= 0}
,
\;\;\;\;\;
< r^2 >_{V(P)}^\pi =  - 6 \left. \frac{ d  
\Pi_{V(P)}^\pi}{d Q^2} \right|_{Q^2= 0}
,
\end{eqnarray}

\begin{table}[ht]
\caption{
\small Averaged quadratic radia
for the values of parameters given above
%%  $M_u = M_d = 0.35 $GeV,     
$eB = 0.1 M^2$,
at $|Q| \sim |K| \sim M$.
} 
\centering  
\begin{tabular}{| c | c c  | c c  | } 
\hline\hline  
& $ \sqrt{<r^2>_{S(P)}^n}$ &  $\sqrt{<r^2>_{S(P)}^c}$ 
&
$\sqrt{<r^2>_{V(P)}^n}$ &  $\sqrt{<r^2>_{V(P)}^c} $
  \\
\hline 
& 0.23 \mbox{fm} & 0.20 \mbox{fm} & 0.30 \mbox{fm} & 0.25 \mbox{fm}
\\[1ex] 
\hline  
\end{tabular}
\label{table:Gcurrents}  
\end{table}
\FloatBarrier
The magnetic field induced scalar
a.q.r. is smaller than the magnetic field induced vector a.q.r.
It is important to remind that all the form factors 
are linear in $eB$ in this range of relatively weak magnetic fields.

\begin{figure}[ht!]
\centering
\includegraphics[width=40mm]{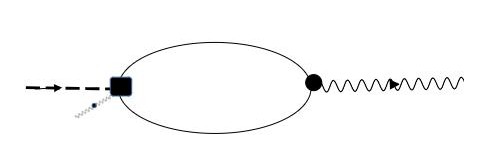}
 \caption{ 
\small
Pion decay to vector quark-antiquark current
 that 
fluctuates into a photon
}
\label{fig:pionphoton}
\end{figure}

An attractive process 
could be the transition of the neutral pion to a photon that could arise because of the vector quark-antiquark current.
This process is shown in Fig. \eqref{fig:pionphoton} and its amplitude can be written as:
\begin{eqnarray}
T_{\pi^0 \to \gamma}^B
&=&
Tr_{D,C,F} \frac{1}{\sqrt{2}}
\sum_{q=u,d}
\int_K
 \gamma_\mu  
 \Pi_{V(P)}^{q,\mu} (K,Q)
\frac{\slashed{K}+\slashed{Q} + M }{
(K+Q)^2 - M^2 }
\gamma^\rho \hat{Q}
\frac{\slashed{K} + M }{
K^2 - M^2 }
\epsilon_\rho (Q).
\end{eqnarray}
where 
 $q$ the charge of a particular up or down quark
of a neutral pion.
This  has been calculated.
However,
it showed to be proportional to $\tilde{Q} \cdot Q$
that is zero.

\FloatBarrier

\section{ Summary and discussion }

We have found unusual vector and scalar single pion couplings to 
constituent quarks induced by relatively weak magnetic fields,
i.e. $eB/{M^2} \simeq   0.1$,
where $M \sim 0.35$ GeV, is a constituent quark, mass that is
$B \sim 6 \times 10^{14}$ T.
Therefore, external magnetic fields can lead to 
changes in pion structure 
according to these
vector and scalar components. 
 The three-point functions involved have been projected in both the Dirac scalar and vector channels for neutral and charged pions.
Although  pseudoscalar pion coupling to constituent quarks have been considered, the axial pion coupling to constituent quarks can also be used. 
Axial pion coupling in  
the scalar three-point function  has been shown to be 
proportional  to the same three-point function calculated with the pseudoscalar coupling. The proportionality factor 
is related to the GTR as shown below Eq. \eqref{Pi-S-A}.
The relation of the vector projection calculated with the pseudoscalar and axial pion couplings is not so simple, and it has not been exploited in this work.

It is interesting to point out that the present calculations of form factors have been carried out based on the
Weinberg's Large Nc EFT,
in spite of the need to change the pion field definition to a linear one in which the pseudoscalar coupling to quarks emerge.
This EFT  therefore offers
a suitable dynamical framework for the calculations of meson form factors that, besides accounting non-perturbative effects by means of the quark and gluon effective propagators, offers a systematic way to calculate corrections to the one-loop level.

The resulting momentum dependent scalar and vector form factors, defined as dimensionless quantities, have been investigated numerically for both neutral and charged pions.
Two kinematical regions have been exploited.
Firstly, by considering 
off-shell pion and on-shell quarks, the relevance for virtual pion exchange type process can be identified.
Second, for off-shell quarks and on-shell massless pions 
the analysis has been focused on processes such as pion emission or absorption.
In the first kinematical region,
the scalar and vector form factors have been found to exhibit different signs. As a consequence, if a pion is exchanged (anomalously)  between a scalar and a vector constituent quark currents, in such a weak magnetic field, it turns out that  a sort of (very weak) repulsive component for the Yukawa-type potential can arise.
Concerning the second kinematical region
(on shell massless pion and off shell quark),
slightly different amplitudes of charged (neutral) pion emission or absorption arise.
They may add or subtract the isospin-breaking difference between neutral and charged pion emission and absorption.
The  neutral and charged pions form factors present slightly differences, mainly in the overall strengths, 
for all channels.
In general, the scalar form factors
($F_{S(P)} \sim 0.05$)
are one order of magnitude smaller than the vector form factors 
($F_{V(P)}  \sim 0.2$)
in both kinematical regions.
However, by multiplying the vector form factor by $\tilde{Q}^\mu$, according to  Eq. \eqref{Fvppi}, they become of the same order of magnitude.
The scalar form factor for
off-shell pion and on-shell quarks is the only one to present negative values, being  all the others positive.
 The same form factors arise for the $\eta(547)$
 and $\eta'(958)$
 with  different numerical coefficients.

Unusual pion couplings to baryons, or conversely constituent quarks, can possibly be used to probe the appearance of magnetic fields in relativistic heavy-ion collisions \cite{PPNP}.
Their effects in hadron structure and dynamics are expected to be possibly identified in   experimental observations.
The strength of the present anomalous pion couplings to constituent quarks is small and signals of their effect would  need high precision measurements. 
Corresponding averaged
quadratic radia (scalar and vector) have been defined
in the usual way and numerical estimations 
were provided.
We   speculated that 
that a transition of a neutral pion into a single photon could take place and it 
could be an indication of magnetic fields.
However, this transition is
forbidden unless
some external interaction modifies the three-momentum of the (possibly) mutating pion.
%This might happen for a 
%neutral pion in a hadron 
%gas endowed with 
%vorticity.
More extensive analysis of possible consequences of the above
 form factors
will be provided elsewhere.

\section*{Acknowledgements}

The authors  thank
O. Savchuk for indicating Refs. \cite{vorticity-particles}.
%F.L.B.
% is a member of the INCT-FNA
%Proc. 464898/2014-5.
F.L.B. acknowledges  partial financial support from
CNPq-307792/2025-0 and CNPq-407162/2023-2. 
M.L. and C. V. acknowledge support from ANID/FONDECYT (Chile) under Grants No. 1220035 and No. 1250206.

\end{document}